%% file: SKA-Supernovae_unmarked-2.tex
\documentclass[a4paper,11pt]{article}
\usepackage{aaskaiid}
\usepackage{orcidlink} 
\usepackage{color}

\newcommand{\ergshz}{erg\,s$^{-1}$\,Hz$^{-1}$}

\title{Supernovae with the Square Kilometre Array}
\ShortTitle{Supernovae with SKA}

\author[1]{Chandra, Poonam
\orcidlink{0000-0002-0844-6563}}
\ShortName{Chandra et al.} % shortened name list for header 
\author[2,3]{Iwata, Yuhei\orcidlink{0000-0002-9255-4742}}
\author[4]{Nayana, A. J.\orcidlink{0000-0002-8070-5400}}
\author[5]{P\'erez-Torres, Miguel\orcidlink{0000-0001-5654-0266}}
\author[6,7]{Rose, Kovi\orcidlink{0000-0002-7329-3209}}
\affiliation[1]{National Radio Astronomy Observatory, 520 Edgemont Rd, Charlottesville VA 22903, USA}
\emailAdd{pchandra@nrao.edu}
\affiliation[2]{Mizusawa VLBI Observatory, National Astronomical Observatory of Japan, 2-12 Hoshigaoka, Mizusawa, Oshu, Iwate 023-0861, Japan}
\affiliation[3]{Astronomical Science Program, Graduate Institute for Advanced Studies, SOKENDAI, 2-21-1 Osawa, Mitaka, Tokyo 181-8588, Japan}
\emailAdd{yuhei.iwata@nao.ac.jp}
\affiliation[4]{Department of Astronomy, University of California, Berkeley, CA 94720-3411, USA}
\emailAdd{nayana@berkeley.edu}
\affiliation[5]{Instituto de Astrof\'isica de Andaluc\'ia (IAA-CSIC), Glorieta de la Astronom\'ia s/n, E-18008 Granada, Spain}
\emailAdd{torres@iaa.es}
\affiliation[6]{Sydney Institute for Astronomy, School of Physics, The University of Sydney, New South Wales 2006, Australia}
\affiliation[7]{Australia Telescope National Facility, CSIRO, Space \& Astronomy, PO Box 76, Epping, NSW 1710, Australia}
\emailAdd{kovi.rose@sydney.edu.au}

\abstract{
This chapter presents the science potential of the Square Kilometre Array (SKA) for studying all classes of supernovae  and their environments. It substantially updates and extends the earlier work of \citet{PerezTorres2015}, originally published in the 2015 Advancing Astrophysics with the SKA (AASKA14) volume, reflecting the dramatic progress in time-domain astronomy and radio instrumentation over the past decade.

We outline how SKA1 and its pathfinders will transform the radio study of core-collapse supernovae (CCSNe)  through sensitive, commensal wide-field surveys capable of discovering hundreds of events per year, providing a dust-unbiased census of massive-star deaths and direct measurements of the volumetric CCSN rate. The same data will probe ejecta–circumstellar-medium (CSM) interaction, shock microphysics, and progenitor mass-loss histories.

Deep, triggered observations of thermonuclear supernovae (SNe Ia)  will allow the SKA to test competing progenitor scenarios by detecting—or definitively excluding—the prompt radio emission expected from single-degenerate systems. The chapter further explores superluminous  supernovae (SLSNe), delayed interaction supernovae   and synergies with facilities such as ALMA, ngVLA, CTA, IceCube-Gen2, and ULTRASAT. Collectively, these studies will turn radio supernova astrophysics from a discovery-limited field into one governed by population statistics.
}

\begin{document}
\maketitle

\section{Introduction}
%{\color{blue}: Yuwei Iwata. Poonam Chandra, Nayana, Miguel, please add your names here...}

Supernovae are among the most energetic explosions in the Universe. In the thermonuclear pathway, a sufficiently massive carbon–oxygen white dwarf in a binary system may accrete material from its companion and approach the Chandrasekhar mass, triggering a runaway nuclear fusion instability and producing a Type Ia supernova \citep{WhelanIben1973, Nomoto1982}. Massive stars, on the other hand, end their lives through the gravitational collapse of their cores, giving rise to the diverse family of CCSNe; \citealt{Bethe1985, Woosley2002}.
Both SNe Ia and CCSNe exhibit several observational 
subclasses, primarily distinguished by their optical 
spectroscopic and photometric properties. Classical SNe Ia 
are characterized by the absence of hydrogen and the 
presence of prominent Si II $\lambda 6355$ 
\AA\, absorption \citep{filippenko1997optical}. A subset of SNe Ia show narrow hydrogen emission features caused by interaction between the ejecta and dense, unshocked H-rich circumstellar material (CSM), forming the class of SNe Ia-CSM \citep{Hamuy2003_2002ic}.
Within the CCSN population, the presence or absence of hydrogen defines Types II and I, respectively. Type II SNe are further separated into IIP and IIL events based on whether their optical light curves exhibit a plateau or a more linear decline. Type I CCSNe are subclassified as Ib or Ic depending on whether helium is detected or not, respectively. Transitional Type IIb supernovae bridge the gap between hydrogen-rich (Type II) and hydrogen-poor (Type I) events. Collectively, the sequence IIP → IIL → IIb → Ib → Ic is interpreted as reflecting progressively greater stripping of the progenitor’s outer layers prior to explosion \citep{filippenko1997optical}. In addition, 
CCSNe that interact strongly with dense CSM are designated with the suffix “n,” producing subclasses such as Types IIn, Ibn, and Icn. For a comprehensive classification overview, we refer the reader to Fig. 1 of \citet{Chandra2025}.

Massive stars, progenitors of supernovae, 
 shed material from their outer layers throughout their evolution, thereby creating a CSM that encases the star. The mass-loss can be driven by radiation, binary interactions, and/or wave-driven processes \citep{Puls2008}. In the most massive stars, the mass-loss rates can exceed those of solar-type stars by several orders of magnitude.
Although such extensive and often episodic mass loss complicates a one-to-one mapping between supernova classes and the zero-age main sequence (ZAMS) masses of their progenitors, the resulting CSM preserves valuable information about the progenitor’s evolutionary history. Consequently, detailed studies of ejecta–CSM interaction (see § \ref{sec:csm}), which is most relevant for SKA freqencies,  offer a powerful means to trace supernova properties back to their progenitor systems and to probe the physical mechanisms governing late-stage stellar mass loss.

  \begin{figure}[h]
\begin{center}
 \includegraphics[width=0.88\textwidth]{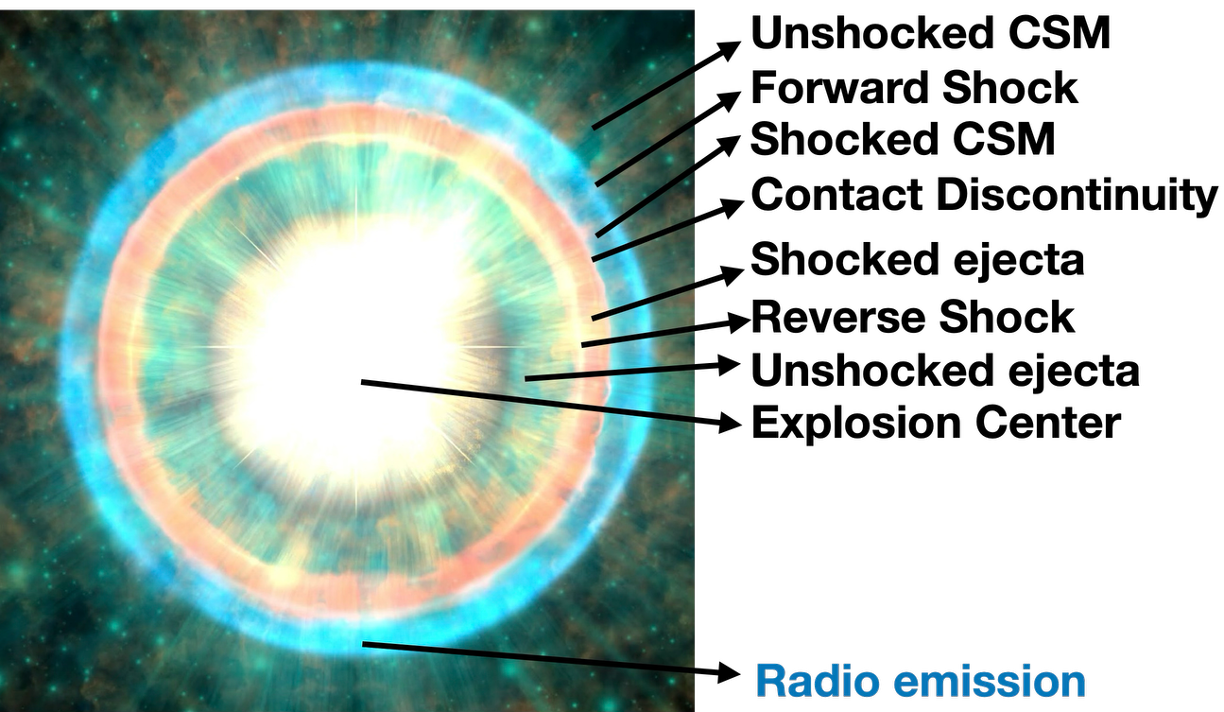} 
 \caption{Circumstellar interaction picture in a typical SN. The  image shows ejecta-CSM interaction and the formation of various  shocked and unshocked regions. The radio emission typically originates from the forward shock.}
   \label{fig:CSM}
\end{center}
\end{figure} 

\section{Circumstellar interaction and radio emission in supernovae}
\label{sec:csm}
%{\color{blue}: Yuwei Iwata. Poonam Chandra, Miguel Pérez-Torres, Nayana, please add your names here...}

When the supersonically expanding SN ejecta encounters the slower-moving CSM, it generates a forward shock that propagates outward into the CSM and a reverse shock that travels back into the expanding ejecta. These two shocks are separated by a contact discontinuity, a region prone to Rayleigh–Taylor and other hydrodynamic instabilities that can amplify magnetic fields (Fig. \ref{fig:CSM}). The shocks heat the plasma to high temperatures and accelerate electrons to relativistic energies; these relativistic electrons, interacting with the amplified magnetic fields, produce non-thermal synchrotron emission that is most prominently observed at radio wavelengths \citep{Chevalier1982a, Chevalier1982b}.
The hot shocked gas also emits thermal and non-thermal X-rays, while X-ray–excited unshocked ejecta or emission from the vicinity of the contact discontinuity can generate additional optical features such as H$\alpha$ emission \citep{Chugai2007}. Among these diagnostics, however, radio emission provides the cleanest and most direct probe of the CSM density structure and, by extension, the progenitor’s mass-loss history.

%Spectral evolution: The radio spectrum evolves from steeply absorbed (optically thick) to a power-law (optically thin) phase.
%Frequency dependence: High frequencies become transparent earlier than low frequencies — hence, multi-frequency monitoring allows us to trace shock evolution.

 Radio emission, which is synchrotron in nature,  is absorbed at early times, either via synchrotron self-absorption (SSA) and/or free–free absorption (FFA) \citep{Chevalier1998}.  
The flux density 
under SSA and FFA can be written,  respectively \citep{Chandra2025}:
\begin{align}
F_{\nu}(t) &= K_{1}\,\nu^{5/2}\,t^{a}\left[1-\exp\!\bigl(-\tau_{\nu}^{\mathrm{SSA}}\bigr)\right],\\ \nonumber
\tau_{\nu}^{\mathrm{SSA}} &= K_{2}\,\nu^{-(p+4)/2}\,t^{-(a+b)},
\end{align}
and 
\begin{align}
F_{\nu}(t) &= K_{1}\,\nu^{-\alpha}\,t^{-\beta}\,
             \exp\!\bigl[-\tau_{\nu}^{\mathrm{FFA}}(t)\bigr],\\ \nonumber
\tau_{\nu}^{\mathrm{FFA}}(t) &= K_{2}\,\nu^{-2.1}\,t^{-\delta},
\end{align}

Here $K_1$ and $K_2$ are the flux density and absorption normalization parameters, respectively; and $\tau_{\nu}^{\mathrm{SSA}}$ and $\tau_{\nu}^{\mathrm{SSA}}$
are SSA and FFA optical depths, respectively. Parameters  $a$, $b$,
$\alpha$, $\beta$, $\delta$ are power law indices, with $p$
being the electron energy index,   related to $\alpha$ as $\alpha=(p-1)/2$ and $\delta$ upon ejecta density structure, $\rho_{\rm ej} \propto r^{-n}$,  as $\delta=(n-3)/(n-2)$.
Thus radio emission typically rises as the expanding shock becomes optically thin to its absorption/s, peaks, and then decays. The radio spectrum evolves from steeply absorbed (optically thick) to a power-law (optically thin) phase. As high frequencies become transparent earlier than low frequencies, hence, multi-frequency monitoring allows us to trace shock evolution and yields %shock radius/velocity, magnetic field, and density profile. 
important parameters. Radio observations map the CSM density profile and time variability, and enable us to constrain shock physics, magnetic fields, forward shock velocity, and shock microphysics.
High cadence early- to late-time radio observations also reveal light-curve bumps and re-brightening events, thought to be caused by shock interaction with discrete shells or clumps of CSM \citep[e.g.;][]{Anderson2017,Rose2024}.

SKA is an ideal telescope for SN studies owing to its unprecedented sensitivity and large field of view (FoV). The SKA capabilities will allow us to conduct both targeted follow-up observations and wide-field, high-cadence surveys via commensal use. The SKA will be essential in probing SN-CSM interaction at early to late times to constrain their progenitors. Furthermore, its capability for unbiased wide-area surveys makes the SKA a suitable instrument to discover hard-to-find supernovae in dusty regions where optical surveys fail. In the next several sections, we lay out the topics which will be important to study with SKA in this field.

\section{Radio emission from thermonuclear supernovae}

\subsection{Constraining progenitor models}
%{\color{blue}: Miguel Pérez-Torres, Poonam Chandra. Please add your names here...}

Despite the ubiquity of SNe Ia as cosmological distance indicators, the nature of their progenitor systems remains unsettled.  The two main paradigms---the single-degenerate (SD) channel, in which a carbon--oxygen white dwarf accretes material from a non-degenerate companion until reaching the Chandrasekhar mass, and the double-degenerate (DD) channel, involving the merger of two white dwarfs---predict fundamentally different circumstellar environments \citep{Maoz2014}.  SD systems should exhibit detectable CSM and thus prompt radio and X-ray emission from ejecta--CSM interaction, whereas DD systems should be radio-silent at early times.  Radio non-detections therefore provide direct upper limits on pre-explosion mass loss and can, in turn, rule out entire regions of SD parameter space \citep{PerezTorres2014}.

SNe Ia have eluded radio detections until recently. 
Deep VLA and eMERLIN observations of nearby events such as SNe~2011fe and~2014J yielded 3$\sigma$ upper limits of $\dot{M}\lesssim(6$--$7)\times10^{-10}\,(v_{\mathrm{w}}/100~\mathrm{km~s^{-1}})\,M_\odot~\mathrm{yr^{-1}}$ \citep{PerezTorres2014}.  These results seem to exclude symbiotic systems, optically thick accretion winds, and most recurrent-nova configurations as viable progenitors.  They leave only very tenuous or quiescent systems consistent with observations, strengthening the case for DD progenitors in at least a fraction of SNe~Ia.

The Ia-CSM SN~2020eyj was the first and the only SN Ia of any subtype to have shown radio emission \citep{Kool2023}. The radio emission was detected on 605 and 741 days, post-optical discovery, with 5 GHz radio spectral luminosity in the range $1.2$--$1.6\times10^{27}$\,\ergshz{}.
The lack of radio emission from other SNe Ia-CSM is therefore intriguing, despite detections at other wavelengths indicating presence of dense CSM. One reason could be excessive absorption which can lead to non-detectable radio emission due to progressively decreasing synchrotron strength with time. 
 Another possibility is different microphysics of shocks in thermonuclear SNe compared to that in CCSNe, which may alter the efficiency of synchrotron production. 
SKA radio limits and detections of SNe Ia combined with SKA VLBI will be able to probe the parameter space of the microscopic physics.

\subsection{SKA capabilities and Observing strategy of SNe Ia}
The sub-$\mu$Jy sensitivity of SKA1-Mid (AA4 configuration; $\sigma_{1\,\mathrm{h}}\simeq0.7~\mu$Jy at 1.7~GHz) will push these limits one order of magnitude deeper.  A non-detection at this level for an SN~Ia at the distance of M\,82 ($D\simeq3.5$~Mpc) would constrain the wind-density parameter to $\dot{M}/v_{\mathrm{w}}\lesssim5\times10^{-13}~M_\odot~\mathrm{yr^{-1}}~\mathrm{km~s^{-1}}$, effectively excluding all single-degenerate channels with hydrogen-rich companions.   Conversely, a detection of early radio emission would  provide the physical parameters and, when combined with the optical and X-ray data, directly measure the density, composition, and microphysical parameters of the surrounding medium, providing the first definitive identification of an SD progenitor.   The full SKA will reach $\sim70$~nJy~h$^{-1/2}$, enabling detections or limits on prompt emission out to $\sim25$~Mpc, where $\sim2$ SNe~Ia per year are expected, yielding a statistically meaningful sample within a decade of operation, sufficient to carry out population synthesis studies of SNe Ia.

%\paragraph{Late-time monitoring.}
In the double-degenerate scenario, merger ejecta expanding into the tenuous interstellar medium may yield slowly rising radio emission on decade timescales as the shock sweeps up ambient material \citep{2016ApJ...823..100H, 2025ApJ...995...54G}.  Regular SKA monitoring of decades-old SNe~Ia will therefore test this prediction and may reveal the delayed onset of synchrotron radiation characteristic of merger-driven explosions.

%\paragraph{Observing strategy}

 For Type Ia SNe, the ideal strategy is to trigger within 1--3 days of explosion, in order to probe the very compact CSM structure. For example SN 2011fe was observed in radio beginning about 2 days after explosion, and those data set definitive limits on the immediate environment \citep{2012ApJ...746...21H}. A practical approach of  triggering  nearby and very young SNe Ia is likely to result in best constraints. Despite radio non-detection at early times, if the optical spectra shows presence of interaction or 
time-variable narrow absorption, then additional efforts need to be made to monitor it at later times, say until a year.

%Thus a  practical programme would combine rapid target-of-opportunity (ToO) triggers for all SNe~Ia within $\lesssim30$~Mpc, beginning within a 1--3 days after optical discovery, with yearly deep follow-up of historical events. 
%The total time demand is modest ($\lesssim$20~hr~yr$^{-1}$) yet decisive for distinguishing progenitor channels.

\section{Radio emission from Core-collapse supernovae}

\subsection{Clues to the progenitor}

\begin{figure}[h]
\begin{center}
 \includegraphics[width=0.90\textwidth]{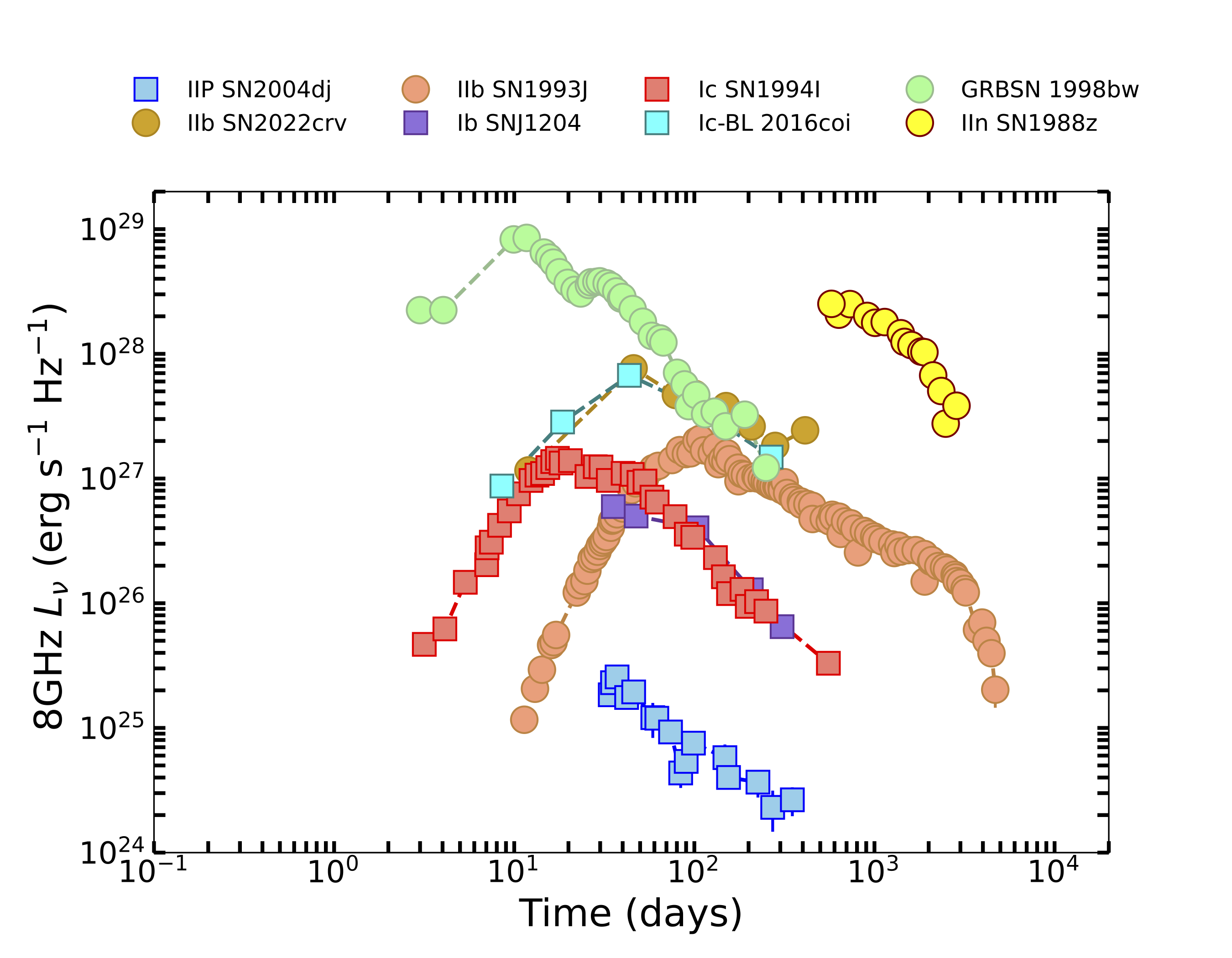} 
 \caption{Plot of 8 GHz radio light curve for various types of CCSNe. The diversity in the radio luminosity, time to peak and extent of the radio emission is apparent in different classes and indicate towards diversity in progenitors. The data are taken from \citet{Nayana2018, Weiler2007, Weiler2011, KUlkarni1998, Gangopadhyay2023, Chandra2019, Nayana2020}.}
\label{fig:radio}
\end{center}
\end{figure} 

Massive stars can experience multiple channels of mass loss during their lifetimes. The dominant mechanisms include radiation-driven winds, binary interaction–induced stripping, and wave-driven mass loss \citep{Puls2008, Smith2014}. Because mass loss shapes the CSM, different progenitor systems produce markedly different CSM structures and densities. Studying the CSM across various spatial and temporal scales provides crucial insight into the nature and evolution of SN progenitors \citep{Chandra2025}.
Radio light curves and spectra are particularly powerful probes of this connection. The epoch of radio turn-on, the nature of absorption, the timing and strength of the radio peak, and the overall extent of emission each encode distinct physical processes associated with specific progenitor environments (Fig. \ref{fig:radio}). Historically, most radio studies have concentrated on higher frequencies, where sensitivity to the evolving CSM density profile is reduced. Low-frequency observations, however, allow the emission to be followed over longer timescales, thereby revealing a more extended record of the progenitor’s mass-loss history. This is where SKA stands. 

Since supernova shock velocities are typically $100$--$1000$ times greater than those of the CSM winds \citep{Chevalier2003}, the expanding shock effectively samples material ejected hundreds to thousands of years before explosion. This makes ejecta-CSM interaction an invaluable “time machine” for reconstructing the pre-SN evolution and mass-loss history of massive stars during their advanced nuclear-burning phases. 
The pronounced observational diversity in radio bands among supernovae exhibiting ejecta-CSM interaction underscores the complex mass-loss histories of their progenitors, involving multiple channels such as steady winds, binary mass transfer, and eruptive pre-SN outbursts. The ejecta-CSM interaction remains the most effective diagnostic of progenitor mass-loss rates and CSM densities over timescales ranging from the final year to several centuries before explosion.

\subsection{Confined CSM, shock breakout}
%{\color{blue}: Yuwei Iwata, Nayana}

Recent progress in optical wide-field transient surveys has enabled the discovery of supernovae within hours of explosion. Rapid optical follow-up spectroscopy has revealed transient, high-ionization narrow emission lines emerging shortly after the explosion \citep[e.g.,][]{Yaron2017}. These so-called “flash” features indicate the presence of a denser CSM at small radii (i.e., confined CSM) located very close to the progenitor, typically within $\lesssim 10^{15}\,{\rm cm}$. Such confined CSM implies that the progenitor star experienced an episode of enhanced mass-loss during the final years to decades before core collapse, reaching the mass-loss rates of $\gtrsim 10^{-3}\, M_{\odot}\,{\rm yr^{-1}}$, which is several orders of magnitude higher than the steady wind of a typical red supergiant ($\sim 10^{-6}\, M_{\odot}\,{\rm yr^{-1}}$) \citep[e.g.,][]{Goldman2017}. Statistical studies further suggest that a significant fraction of Type II SNe show evidence for such confined CSM \citep[e.g.,][]{Forster2018}.

While optical “flash” spectroscopy probes the innermost region, radio observations trace the interaction with the more extended CSM. However, the confined CSM can strongly absorb the synchrotron emission via free–free absorption at early times, particularly at lower frequencies, where the optical depth is higher. This often results in non-detections in the low-frequency bands during the first few days, or in some cases up to hundreds of days, after the explosion. Such early non-detections themselves provide valuable constraints on the optical depth and density of the confined CSM. As the shock expands and the medium becomes optically thin, radio emission emerges, revealing the density structure beyond the confined region. Therefore, high-cadence, multi-frequency, long-term radio monitoring -- from the earliest phases  at higher frequencies with SKA-Mid through to the later evolution at lower frequencies, employing SKA-Low  is essential to capture the transition from absorption-dominated to emission-dominated phases, and to map the density variations that record the progenitor’s final mass-loss history. In § \ref{sec:ccsn}, we expand on observing strategies.

%\subsection{gamma rays from %supernovae}
%%{\color{blue}: Yuwei Iwata}
%Merged into Section 10?

\subsection{Supernovae in dense CSM: secondary leptons}
\label{subsec:secondary}

\begin{figure}[h]
\begin{center}
 \includegraphics[width=0.80\textwidth]{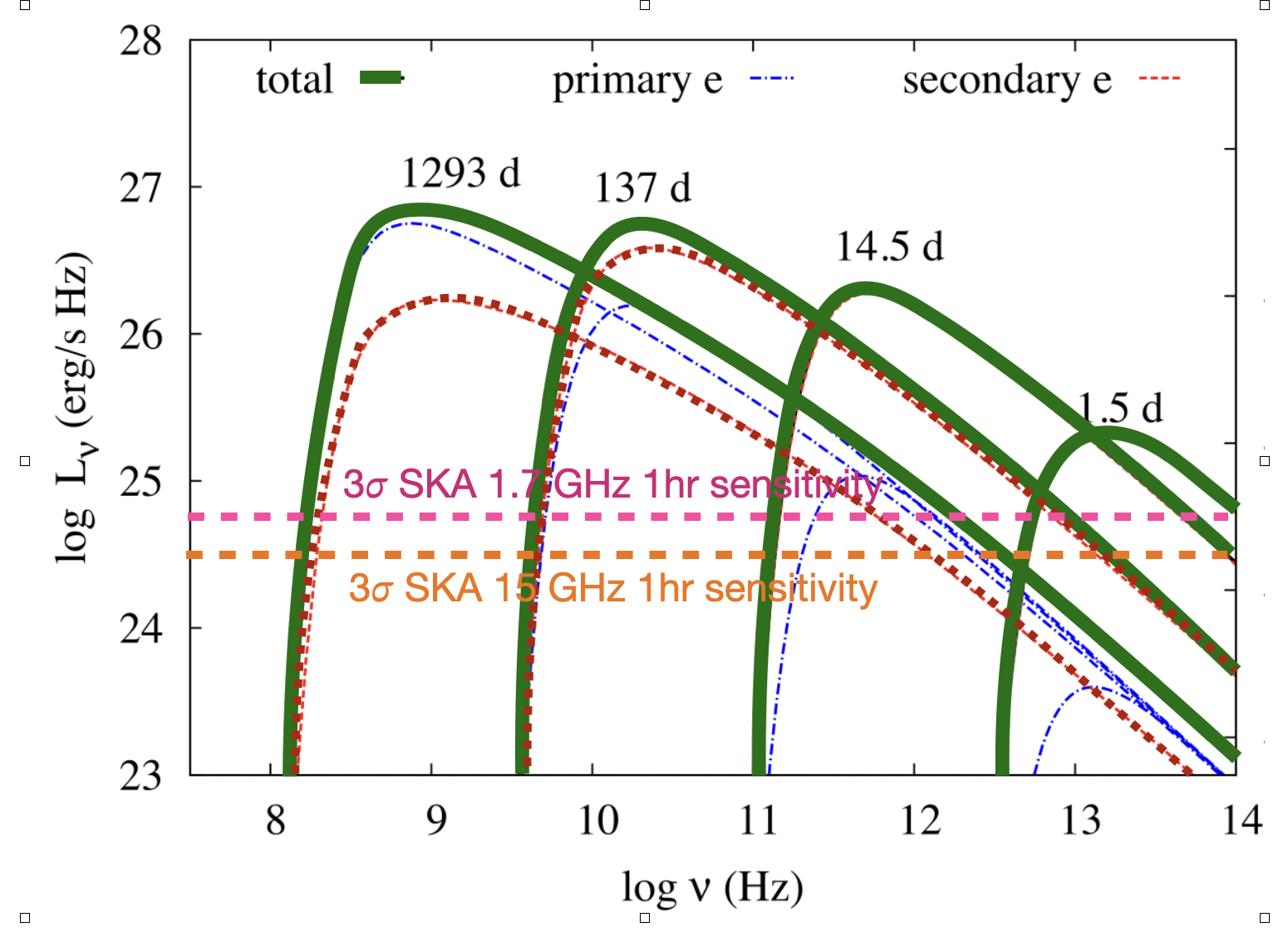} 
 \caption{The contribution of the primary (blue dash–dotted lines) and secondary (red
dashed lines) electrons to the total synchrotron emission  (green solid line) for a wind density  CSM for parameters mentioned in \citet{Petropoulou2016},  the paper from which this figure is adapted and a distance of 40 Mpc is assumed. We also show 3-$\sigma$ SKA-Mid sensitivities at 1.7 GHz and15 GHz at 40 Mpc. }
\label{fig:secondary}
\end{center}
\end{figure} 

%{\color{blue}: Yuwei Iwata, Nayana}
There are growing pieces of evidence for dense  CSM shells around CCSNe. The collision of the SN shock with such dense environments results in efficient conversion of shock kinetic energy to radiation which has been proposed as one possible powering source of superluminous supernovae \citep[SLSNe;][and see Section \ref{sec:slsn}]{Falk1973,Ofek2007,Smith2007,Quimby2011}. These interaction-powered supernovae can act as efficient cosmic ray accelerators for a range of CSM parameters \citep{Murase2011,Katz2012,Kashiyama2013}.
 The accelerated protons undergo $pp$ collisions and experience pion losses that will produce gamma-rays, neutrinos, and secondary leptons.  
  While primary electrons are
injected at shock, the 
secondary electrons are
injected throughout shocked region and can emit detectable synchrotron radiation.
 The actual synchrotron spectrum and flux densities depend on various parameters including ejecta velocity and mass, and physical location of the CSM shell with respect to the progenitor star. The emission could peak with flux density $\approx 0.01$--$1$\,mJy at $3$--$3000$\,GHz for an SN like SN~2006jd at distances of hundreds of Mpc \citep{Murase2014}.  In Fig. \ref{fig:secondary}, we show radio light curves taken from \citet{Petropoulou2016},  decomposed into contributions from
primary and 
secondary electrons and their relative importance with time.
Secondary electrons dominate early radio emission, but their contribution declines faster than primaries, leading to a break/flattening in the light curve. There is no confirmed detection of synchrotron emission from secondary leptons from supernovae as of yet. \citet{Nayana2025} reported millimeter emission from SN\,2023ixf in excess to the standard synchrotron spectrum and attributed it to possible emission from secondary leptons.  The sensitivity of current radio arrays is a limiting factor in detecting this emission, and the SKA era, mainly SKA-Mid combined with upgraded ALMA, will make significant progress towards detecting/placing robust limits on synchrotron emission from secondary leptons.

%\subsection{Supernovae surveys in radio bands: an unbiased search}

\begin{figure}[h]
\begin{center}
 \includegraphics[width=1.0\textwidth]{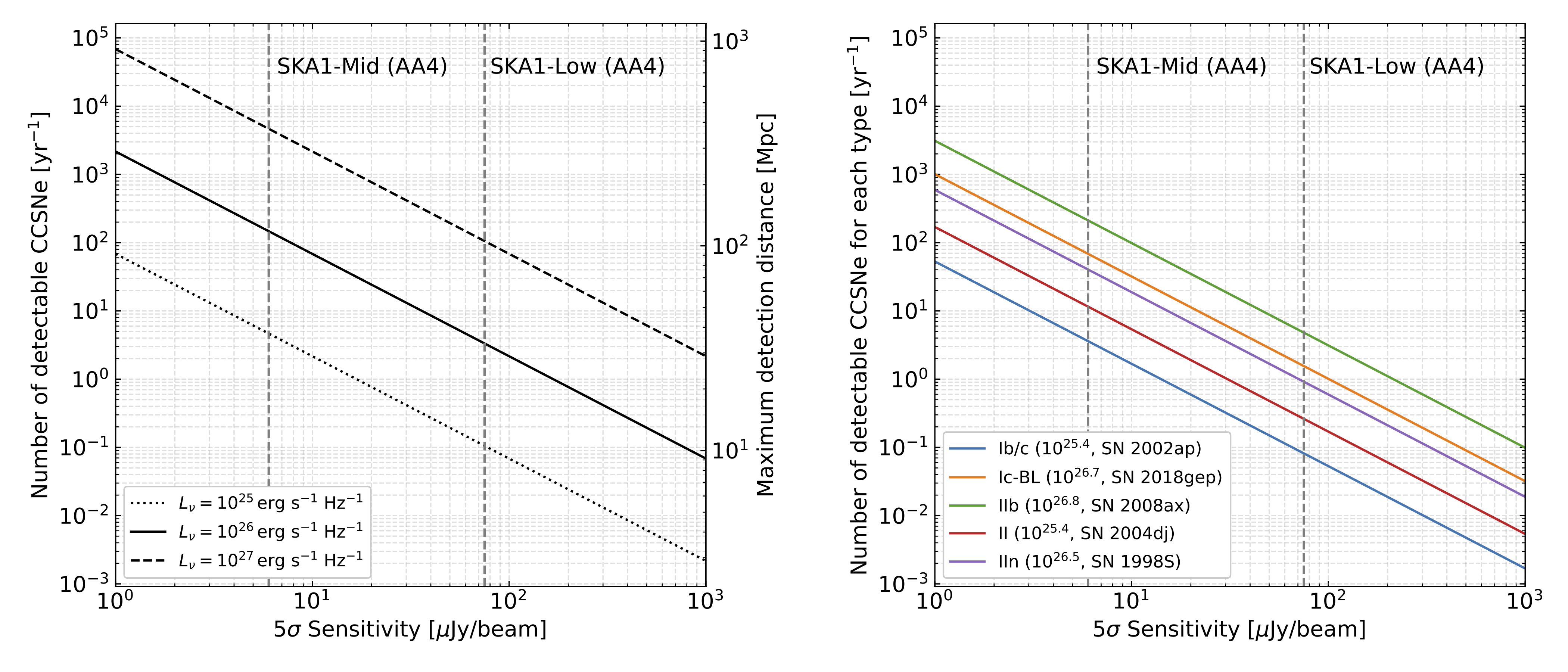} 
\caption{(\textit{Left})  Estimated number of detectable CCSNe per year and the maximum observable distance. Three cases of the peak spectral luminosity are shown: $L_{\nu}$ = $10^{25}$, $10^{26}$, and $10^{27}$ \ergshz. Vertical lines indicate the assumed 1-hour (5$\sigma$) sensitivities of SKA1-Low and SKA1-Mid for the AA4 configuration. (\textit{Right}) Same as the left panel, but divided by SN type. The assumed SN fraction of each type is taken from \citet{Ma2025}.
The average peak luminosities (from \citealt{Bietenholz2021}) and representative SN events corresponding to those luminosities (from \citealt{PerezTorres2015}) are shown in the labels. Both panels assume a CCSN rate of $N \approx 10 (D/30\,{\rm Mpc})^3$ yr$^{-1}$ and a yearly survey area of 10,000 deg$^2$.}
\label{fig:Number}
\end{center}
\end{figure} 

\subsection{Late time radio emission from core-collapse supernovae}

Radio emission from CCSNe is typically monitored for $\approx 1$--$2$ years except for a few very nearby events. The radiation is synchrotron in origin and displays high brightness temperatures, and is often attenuated by free-free absorption by the ionized medium in the line of sight and/or synchrotron self-absorption by the same plasma. As a result of the temporal evolution of combined emission and absorption processes, the radio light curves trace a general pattern of a steep rise to a maximum flux density ($F_{\rm pk}$) followed by a power-law decay ($F_{\nu} \propto t^{\beta}$) with decay index of range $\beta \approx -1\rm \,to\,-3$. The time to peak radio emission ($t_{\rm pk}$) ranges from a few days to several years, and the corresponding peak spectral luminosities ($L_{\rm pk}$) cover $\sim 10^{23}$–$10^{29},{\rm erg,s^{-1},Hz^{-1}}$ \citep{Bietenholz2021}, with both quantities exhibiting large variation across different SN types. The follow-up observations of individual supernovae are usually carried out to trace the evolution of this radio emission and as the light curve decays, the emission is expected to become undetectable with the current generation radio telescopes.

  However, late time studies have thrown some surprises. \citet{Stroh2021} reported late-time ($1$--$60$ yrs post explosion) radio emission from 19 CCSNe by cross-matching optical positions of known SN with the first data release of the VLA Sky Survey \citep[VLASS;][]{Lacy2020}. The luminosities of this sample spans $L_{\rm pk} \approx 10^{26}$--$10^{29}\, \rm erg\,s^{-1}\,Hz^{-1}$, reaching the peak luminosities of early radio light curves.
Similar work was conducted by \cite{Rose2024}, who found radio emission from 29 archival SNe using the Variables and Slow Transients \citep[VAST;][]{Murphy2013,Murphy2021} pilot survey conducted with the SKA precursor Australian SKA Pathfinder telescope \citep[ASKAP;][]{Hotan2021}. The VAST detections covered a similar range of luminosities to the VLASS sample and identified five late-time radio re-brightening events.
Broadly speaking, there are three possible scenarios to explain the late-time bright radio emission -- (i) SN shock interaction with a dense CSM shell possibly ejected at an earlier phase of mass-loss in the progenitors life \citep{Chevalier2006}, (ii) Emission from relativistic jets launched at the time of explosion which was initially off-axis and coming to the line of sight as the jet decelerates \citep{Granot2002}, and (iii) Emission from the central compact object left behind by the core-collapse like a newly born pulsar wind nebula \citep{Slane2017}. The sample of CCSNe with late-time radio emission offers an unique opportunity to explore these different scenarios. 

While the VLASS and VAST samples cover opposite celestial hemispheres and are limited by an rms noise of $\approx 120\, \mu\rm Jy$ and $\approx 250\, \mu\rm Jy$, respectively, SKA's unique sensitivity and coverage will unravel a significantly bigger sample of these late-time radio emitters. Low-frequency is particularly interesting as the emission component from the initial blast wave might be still bright in addition to the late-time component, as evidenced by \citep{Rose2024}. 
With one hour 5$\sigma$ sensitivities of $6\, \mu\rm Jy/beam$ (SKA1-Mid) and $75\, \mu\rm Jy/beam$, (SKA-Low) in AA4, we can extrapolate the \cite{Rose2024} VAST scaled estimate to expect  $10^3$--$10^4$ detectable re-brightening events and each year with the SKA.
A systematic study of such a statistically significant sample will answer questions about the off-axis jets in relativistic supernovae (Type Ic-BL), complex mass-loss history of progenitors, and emission from the youngest compact remnants. 
%Moreover, follow-up observations with SKA-VLBI for sufficiently bright events would allow spatially resolving their expanding or jetted structures and measuring their expansion dynamics with high fidelity. 
SKA, when used in VLBI mode, will be able to reach sensitivity 0.5--1 $\mu$Jy in a few hours, and provide mas resolution \citep{Rioja01.2026.SKA}
.  
%SKA-VLBI  will reach $\mu$as resolution. 
%While the SKA-VLBI detectability depends upon supernova type and luminosity, SKA-VLBI can detect  supernovae out to hundreds of Mpc.
%Also, SKA-VLBI will resolve CCSNe  within at least 30 Mpc, yet we will able to derive meaningful size constraints for supernovae within $\sim$ 100 Mpc.
We note here that the angular resolution of SKA-VLBI will not be improved beyond what is achievable with current facilities, hence we realistically expect to resolve CCSNe out to a distance of $\sim$30 Mpc. Furthermore, the better sensitivity of the SKA1 (see Fig.~\ref{fig:Number}) will allow the derivation of meaningful size constraints for recently exploded CCSNe out to about 100 Mpc \citep{TaoAn02.2026.SKA}. 
Given the volumetric CCSN rate of $\sim 7\times10^{-5}$ yr$^{-1}$ Mpc$^{-3}$ \citep{Pessi2025}, the potential number of targets to be resolved per year is about 8 supernovae, which is a sensible number of targets for specific follow-up with SKA-VLBI within 30 Mpc, and amounts to $\sim$280 within 100 Mpc, which leaves room to select the brightest and most relevant targets. 
We note here that the limiting issue is the time needed to resolve the expanding shell of a CCSN. Indeed, at least 15 yr would be needed to barely resolve a CCSN at 100 Mpc (at 15 GHz). For this reason, targeted SKA-VLBI observations should be done only on the closest and brightest targets of each subtype. In addition, SKA-Mid and SKA-Low, in VLBI fashion, will allow us to image old exploded CCSNe, which current VLBI arrays cannot resolve out due to sensitivity limitations.

\subsection{SKA capabilities and observing strategies for CCSNe}
\label{sec:ccsn}
%{\color{blue}: Miguel Pérez-Torres, Poonam Chandra}

 Untargeted and commensal surveys by SKA will find hundreds of CCSNe per year with SKA-1 and an order of magnitude more with SKA-2, including heavily obscured events in dusty environments that optical surveys miss. Such surveys will help resolve the ``missing SNe'' problem and provide robust measurements of the volumetric CCSN rate and the subtype 
 distribution. Fig. \ref{fig:Number} shows the estimated number of detectable CCSNe per year. If we assume a peak luminosity of $10^{26}$ 
 \ergshz, SKA-1 Mid would detect $\sim 100$ SNe yr$^{-1}$ within a maximum distance of $\sim 100$ Mpc. While the majority of these detections will rely on cross-matching with optically discovered SNe, image-plane slow 
 transient searches across multiple epochs could also yield a population of radio-only detected SNe. Even without optical counterparts, these radio transients can potentially be identified as SNe if their peak radio 
 luminosities and the shape of the light curve (or the characteristic evolutionary timescales) can be adequately constrained \citep[e.g.,][]{Brunthaler2009}. This approach opens a pathway to construct an unbiased sample of radio SNe. When estimating the number of detectable supernovae by subclass, using the average peak luminosity \citep{Bietenholz2021} and 
 relative occurrence rate for each type \citep{Ma2025}, Type IIb events are predicted to be the most numerous, with approximately hundreds of sources expected to be detectable per year (see right panel of Fig. \ref{fig:Number}). These estimates depend on several uncertain parameters and particularly on the peak luminosity: for subclasses whose peak luminosities are close to the detection threshold, detection is feasible only if the source is observed near its peak. On the other hand, low-frequency light curves evolve over much longer timescales, which increases the likelihood that sources can be detected.

 For CCSNe the trigger strategy depends upon not just how young it is, but the subtype of a supernova. A large compilation of radio light curves has indicated that the characteristic peak times of roughly $10^{1.1}$--$10^{1.5}$ days for SNe Ib/Ic to SNe IIb, $10^{1.6}$ days for non-IIn SNe II and $10^{3}$ days for SNe IIn \citep{Bietenholz2021}. 
Hence the optimum trigger strategy is to trigger SNe IIb, Ib and Ic within 2--7 days, SNe IIP within 5--10 days and SNe IIn from a couple of weeks to a month.
The follow up should be longer for SNe showing signs of structured CSM, strong H$\alpha$, luminous X-rays, or optical bumps/rebrightenings.
A systematic compilations of detected supernovae and statistical study of their peak luminosities will improve correlation between typical radio peak luminosities and rise times across CCSNe subtypes and emphasize diversity \citep{Bietenholz2021}.
To this end, a near ideal practical strategy is to carry out an untargeted survey with SKA-Low (wide-field) and SKA-Mid (high-sensitivity), and then multi-band follow-up with 3--5 SKA bands. A logarithmic cadence will be crucial, given the power-law time evolution of SNe \citep[e.g.,][]{Weiler2010}. In addition, follow-up observations with other radio facilities will provide essential complementary coverage, both by accessing frequency ranges not available to the SKA and by enabling rapid, flexible scheduling \citep[e.g.,][]{Iwata2025}.

\section{Superluminous supernovae with the SKA}
\label{sec:slsn}
%{\color{blue}: Please add your names here...}

Superluminous supernovae (SLSNe) represent a relatively recent, observationally defined class of explosions whose peak luminosities exceed those of canonical supernovae by more than an order of magnitude \citep{GalYam2012, Galyam2019, Moriya2024}. Although SLSNe may arise from various stellar death pathways, their extreme luminosities are thought to be powered by one or more of three primary mechanisms: (i) radiation from CSM interaction, (ii) energy injection from a central engine such as magnetar spin-down or fallback accretion, or (iii) radioactive decay associated with the pair-instability mechanism \citep{Galyam2019, Moriya2024}.
To date, only two SLSNe—both hydrogen-poor SLSNe-I—have been detected in the radio: PTF10hgi \citep{Eftekhari2019, Eftekhari2021, Mondal2020} and SN 2017ens \citep{Margutti2023}. In PTF10hgi, radio emission was discovered approximately seven years after explosion; modeling indicated that the event was magnetar-powered, with the radio emission arising from a pulsar wind nebula confined by the surrounding CSM. In contrast, SN 2017ens exhibited Balmer emission lines in its optical spectra about 100 days post-explosion \citep{Chen2018}, and its radio detection nearly three years later revealed strong ejecta-CSM interaction as the dominant power source.
Because SLSNe are typically found at large cosmological distances—and their early-time radio emission is often absorbed—the steadily declining synchrotron flux makes them challenging to detect once the optical depth to absorption drops below unity at late epochs and low frequencies. SKA will prove to be a diagnostic machine that can separate the powering channels as the each mechanism is likely to imprint a distinct radio timescale and evolution. While magnetar powered SLSNe are expected to show a slow rise and evolution, the off-axis jet will show a sharper rise on months to years timescales followed by the decay. If CSM interaction is powering SLSNe, then it is not possible to have a smoothly decaying CSM and one is likely to see bumps and wiggles in the radio lightcurve. The enhanced sensitivity of the SKA will be critical for uncovering and characterizing such faint, late-time radio counterparts.

%\section{High-z and gravitationally lensed supernovae with the SKA}
%{\color{blue}: Please add your names here...}

%\section{SN-SNR connection and clues from long wavelengths}
%{\color{blue}: A. J. Nayana, Miguel Pérez-Torres, Please add your names here...}

\section{Multiwavelength, multi-messenger synergies with SKA}
%{\color{blue}: Yuhei, Nayana, Poonam}

With the advent of next-generation facilities, multiwavelength and multimessenger studies of supernovae will enter a transformative era. The launch of ULTRASAT in 2027 will provide an unprecedented wide field of view and rapid public alerts \citep{BenAmi2022}. ULTRASAT will routinely capture the earliest ultraviolet signatures of stellar explosions, including shock breakout and flash-ionization phases. When combined with SKA observations of early-time radio emission—particularly the radio turn-on that encodes the circumstellar density structure—joint modeling will enable us to break long-standing degeneracies in ejecta–CSM interaction models.
The SKA will operate synergistically with both ALMA (upgraded) and the ngVLA, together covering a broad swath of the radio-to-millimeter spectrum. ngVLA frequencies will be highly complementary to those of the SKA, providing wide-band spectral coverage that is crucial for diagnosing cooling processes and the evolving microphysics \citep{Kadler02.2026.SKA}. ALMA’s millimeter observations will probe the high-frequency synchrotron tail as well as thermal free–free and dust emission, while SKA measurements at centimeter wavelengths will tightly constrain absorption processes. Taken together, ALMA, ngVLA, and SKA will allow us to disentangle contributions from primary and secondary electrons in dense CSM environments \citep{Petropoulou2016}.

 Nearby core-collapse supernovae are also expected to be sources of high-energy ($\gtrsim$ TeV) gamma rays and neutrinos, neither of which has been detected yet (unlike the MeV neutrinos observed from SN 1987A). As explained in \$ \ref{subsec:secondary}, hadronic collisions in the shock–CSM interaction region accelerate protons to relativistic energies, generating neutral and charged pions whose decay produces gamma rays and neutrinos, respectively. In the SKA era, the Cherenkov Telescope Array (CTA) will be fully operational, and the next-generation neutrino facility IceCube-Gen2 is expected to commence observations, offering the exciting possibility of coordinated multimessenger detections of supernova explosions.

%The radio emission from CCSNe primarily originates from synchrotron radiation produced by shock-accelerated electrons in the CSM. In addition, pion decay generates secondary electrons that can contribute to the observed radio flux densities. In the case of the most recent nearby CCSN, SN 2023ixf, its radio spectrum shows an excess at high frequencies ($\gtrsim 80$ GHz), which has been interpreted as emission from such secondary electrons \citep{Nayana2025}. These results highlight the importance of obtaining broadband radio spectra to disentangle the primary and secondary electron contributions. The SKA, with its high sensitivity at low frequencies, will be a crucial instrument for this purpose. In particular, SKA’s sensitivity will enable detections or provide strong upper limits, even during the early phases after explosion, when free–free absorption significantly attenuates the emission.

%\subsection{Synergies with numerical simulations: role of SKA}
%{\color{blue}: Add your names here}

%\subsection{A supernova in Milky Way?}

\section{Summary}
%{\color{blue}: Poonam, Nayana}

The coming decade will revolutionize supernova astrophysics through the combined power of wide-field discovery surveys and deep, broadband radio follow-up with the SKA. High-cadence time-domain specific will deliver unprecedented samples of nearby and early-phase explosions, providing timely triggers for SKA observations. SKA’s exceptional sensitivity will transform our ability to measure pre-supernova mass loss across all core-collapse subtypes, uncover episodic or eruptive mass-loss episodes, and probe the poorly understood circumstellar environments of stripped-envelope supernovae—including the possible role of central engines. For thermonuclear SNe Ia, SKA’s stringent non-detections will decisively rule out most single-degenerate progenitor channels, while any detection would provide rare, direct constraints on circumstellar density and progenitor physics. Together with complementary facilities such as ALMA, ngVLA, IceCube-Gen2 etc., SKA will enable a comprehensive, multiwavelength and multimessenger view of stellar death, fundamentally reshaping our understanding of how massive stars evolve, explode, and enrich the Universe.

\section{Acknowledgments}
The National Radio Astronomy Observatory and Green Bank Observatory are facilities of the U.S. National Science Foundation operated under cooperative agreement by Associated Universities, Inc. YI was supported by the Japan Society for the Promotion of Science KAKENHI grant JP23K13151. KR thanks the LSST-DA Data Science Fellowship Program, which is funded by LSST-DA, the Brinson Foundation, the WoodNext Foundation, and the Research Corporation for Science Advancement Foundation; his participation in the program has benefited this work.

\bibliographystyle{abbrvnat-maxbibnames4}
\bibliography{chapter} % if your bibtex file is called example.bib

\end{document}

%% file: chapter.bib
@ARTICLE{Pessi2025,
       author = {{Pessi}, T. and {Desai}, D.~D. and {Prieto}, J.~L. and {Kochanek}, C.~S. and {Shappee}, B.~J. and {Anderson}, J.~P. and {Beacom}, J.~F. and {Dong}, S. and {Stanek}, K.~Z. and {Thompson}, T.~A.},
        title = "{Supernova rates and luminosity functions from ASAS-SN: II. 2014─2017 core-collapse supernovae and their subtypes}",
      journal = {\aap},
         year = 2025,
        month = nov,
       volume = {703},
          eid = {A34},
        pages = {A34},
          doi = {10.1051/0004-6361/202556799},
archivePrefix = {arXiv},
       eprint = {2508.10985},
 primaryClass = {astro-ph.HE},
       adsurl = {https://ui.adsabs.harvard.edu/abs/2025A&A...703A..34P}
}

@incollection{TaoAn02.2026.SKA, author = {Tao An and author2 and author3 and author4 and author5},title = {},year = {2026},publisher = {},note = {arXiv search: Report number AASKAII/TaoAn02},booktitle = {Advancing Astrophysics with the SKA -- II (AASKAII)}}

@incollection{Rioja01.2026.SKA, author = {Maria Rioja and author2 and author3 and author4 and author5},title = {},year = {2026},publisher = {},note = {arXiv search: Report number AASKAII/Rioja01},booktitle = {Advancing Astrophysics with the SKA -- II (AASKAII)}}

@incollection{Kadler02.2026.SKA, author = {Matthias Kadler and author2 and author3 and author4 and author5},title = {},year = {2026},publisher = {},note = {arXiv search: Report number AASKAII/Kadler02},booktitle = {Advancing Astrophysics with the SKA -- II (AASKAII)}}

@ARTICLE{PerezTorres2014,
       author = {{P{\'e}rez-Torres}, M.~A. and {Lundqvist}, P. and {Beswick}, R.~J. and {Bj{\"o}rnsson}, C.~I. and {Muxlow}, T.~W.~B. and {Paragi}, Z. and {Ryder}, S. and {Alberdi}, A. and {Fransson}, C. and {Marcaide}, J.~M. and {Mart{\'\i}-Vidal}, I. and {Ros}, E. and {Argo}, M.~K. and {Guirado}, J.~C.},
        title = "{Constraints on the Progenitor System and the Environs of SN 2014J from Deep Radio Observations}",
      journal = {\apj},
         year = 2014,
        month = sep,
       volume = {792},
       number = {1},
          eid = {38},
        pages = {38},
          doi = {10.1088/0004-637X/792/1/38},
archivePrefix = {arXiv},
       eprint = {1405.4702},
 primaryClass = {astro-ph.SR},
       adsurl = {https://ui.adsabs.harvard.edu/abs/2014ApJ...792...38P}
}

@inproceedings{PerezTorres2015,
  author = {Pérez-Torres, M. A. and Alberdi, A. and Beswick, R. J. and Lundqvist, P. and others},
  title = {Core-collapse and Type Ia supernovae with the SKA},
  booktitle = {Advancing Astrophysics with the Square Kilometre Array (AASKA14)},
  year = {2015},
  pages = {60},
  url = {https://pos.sissa.it/215/060},
}

@ARTICLE{Bethe1985,
       author = {{Bethe}, H.~A. and {Wilson}, J.~R.},
        title = "{Revival of a stalled supernova shock by neutrino heating}",
      journal = {\apj},
         year = 1985,
        month = aug,
       volume = {295},
        pages = {14-23},
          doi = {10.1086/163343},
       adsurl = {https://ui.adsabs.harvard.edu/abs/1985ApJ...295...14B}
}

@article{Chevalier1982a,
       author = {{Chevalier}, R.~A.},
        title = "{Self-similar solutions for the interaction of stellar ejecta with an external medium.}",
      journal = {\apj},
         year = 1982,
        month = jul,
       volume = {258},
        pages = {790-797},
          doi = {10.1086/160126},
       adsurl = {https://ui.adsabs.harvard.edu/abs/1982ApJ...258..790C}
}

@ARTICLE{GalYam2012,
       author = {{Gal-Yam}, Avishay},
        title = "{Luminous Supernovae}",
      journal = {Science},
         year = 2012,
        month = aug,
       volume = {337},
       number = {6097},
        pages = {927},
          doi = {10.1126/science.1203601},
archivePrefix = {arXiv},
       eprint = {1208.3217},
 primaryClass = {astro-ph.CO},
       adsurl = {https://ui.adsabs.harvard.edu/abs/2012Sci...337..927G}
}

@ARTICLE{Galyam2019,
       author = {{Gal-Yam}, Avishay},
        title = "{The Most Luminous Supernovae}",
      journal = {\araa},
         year = 2019,
        month = aug,
       volume = {57},
        pages = {305-333},
          doi = {10.1146/annurev-astro-081817-051819},
archivePrefix = {arXiv},
       eprint = {1812.01428},
 primaryClass = {astro-ph.HE},
       adsurl = {https://ui.adsabs.harvard.edu/abs/2019ARA&A..57..305G}
}

@ARTICLE{Eftekhari2021,
       author = {{Eftekhari}, T. and {Margalit}, B. and {Omand}, C.~M.~B. and {Berger}, E. and {Blanchard}, P.~K. and {Demorest}, P. and {Metzger}, B.~D. and {Murase}, K. and {Nicholl}, M. and {Villar}, V.~A. and {Williams}, P.~K.~G. and {Alexander}, K.~D. and {Chatterjee}, S. and {Coppejans}, D.~L. and {Cordes}, J.~M. and {Gomez}, S. and {Hosseinzadeh}, G. and {Hsu}, B. and {Kashiyama}, K. and {Margutti}, R. and {Yin}, Y.},
        title = "{Late-time Radio and Millimeter Observations of Superluminous Supernovae and Long Gamma-Ray Bursts: Implications for Central Engines, Fast Radio Bursts, and Obscured Star Formation}",
      journal = {\apj},
         year = 2021,
        month = may,
       volume = {912},
       number = {1},
          eid = {21},
        pages = {21},
          doi = {10.3847/1538-4357/abe9b8},
archivePrefix = {arXiv},
       eprint = {2010.06612},
 primaryClass = {astro-ph.HE},
       adsurl = {https://ui.adsabs.harvard.edu/abs/2021ApJ...912...21E}
}

@ARTICLE{Eftekhari2019,
       author = {{Eftekhari}, T. and {Berger}, E. and {Margalit}, B. and {Blanchard}, P.~K. and {Patton}, L. and {Demorest}, P. and {Williams}, P.~K.~G. and {Chatterjee}, S. and {Cordes}, J.~M. and {Lunnan}, R. and {Metzger}, B.~D. and {Nicholl}, M.},
        title = "{A Radio Source Coincident with the Superluminous Supernova PTF10hgi: Evidence for a Central Engine and an Analog of the Repeating FRB 121102?}",
      journal = {\apjl},
         year = 2019,
        month = may,
       volume = {876},
       number = {1},
          eid = {L10},
        pages = {L10},
          doi = {10.3847/2041-8213/ab18a5},
archivePrefix = {arXiv},
       eprint = {1901.10479},
 primaryClass = {astro-ph.HE},
       adsurl = {https://ui.adsabs.harvard.edu/abs/2019ApJ...876L..10E}
}

@ARTICLE{Chen2018,
       author = {{Chen}, T. -W. and {Inserra}, C. and {Fraser}, M. and {Moriya}, T.~J. and {Schady}, P. and {Schweyer}, T. and {Filippenko}, A.~V. and {Perley}, D.~A. and {Ruiter}, A.~J. and {Seitenzahl}, I. and {Sollerman}, J. and {Taddia}, F. and {Anderson}, J.~P. and {Foley}, R.~J. and {Jerkstrand}, A. and {Ngeow}, C. -C. and {Pan}, Y. -C. and {Pastorello}, A. and {Points}, S. and {Smartt}, S.~J. and {Smith}, K.~W. and {Taubenberger}, S. and {Wiseman}, P. and {Young}, D.~R. and {Benetti}, S. and {Berton}, M. and {Bufano}, F. and {Clark}, P. and {Della Valle}, M. and {Galbany}, L. and {Gal-Yam}, A. and {Gromadzki}, M. and {Guti{\'e}rrez}, C.~P. and {Heinze}, A. and {Kankare}, E. and {Kilpatrick}, C.~D. and {Kuncarayakti}, H. and {Leloudas}, G. and {Lin}, Z. -Y. and {Maguire}, K. and {Mazzali}, P. and {McBrien}, O. and {Prentice}, S.~J. and {Rau}, A. and {Rest}, A. and {Siebert}, M.~R. and {Stalder}, B. and {Tonry}, J.~L. and {Yu}, P. -C.},
        title = "{SN 2017ens: The Metamorphosis of a Luminous Broadlined Type Ic Supernova into an SN IIn}",
      journal = {\apjl},
         year = 2018,
        month = nov,
       volume = {867},
       number = {2},
          eid = {L31},
        pages = {L31},
          doi = {10.3847/2041-8213/aaeb2e},
archivePrefix = {arXiv},
       eprint = {1808.04382},
 primaryClass = {astro-ph.SR},
       adsurl = {https://ui.adsabs.harvard.edu/abs/2018ApJ...867L..31C}
}

@ARTICLE{Weiler2007,
       author = {{Weiler}, Kurt W. and {Williams}, Christopher L. and {Panagia}, Nino and {Stockdale}, Christopher J. and {Kelley}, Matthew T. and {Sramek}, Richard A. and {Van Dyk}, Schuyler D. and {Marcaide}, J.~M.},
        title = "{Long-Term Radio Monitoring of SN 1993J}",
      journal = {\apj},
         year = 2007,
        month = dec,
       volume = {671},
       number = {2},
        pages = {1959-1980},
          doi = {10.1086/523258},
archivePrefix = {arXiv},
       eprint = {0709.1136},
 primaryClass = {astro-ph},
       adsurl = {https://ui.adsabs.harvard.edu/abs/2007ApJ...671.1959W}
}

@ARTICLE{Gangopadhyay2023,
       author = {{Gangopadhyay}, Anjasha and {Maeda}, Keiichi and {Singh}, Avinash and {Nayana}, A.~J. and {Nakaoka}, Tatsuya and {Kawabata}, Koji S. and {Taguchi}, Kenta and {Singh}, Mridweeka and {Chandra}, Poonam and {Ryder}, Stuart D. and {Dastidar}, Raya and {Yamanaka}, Masayuki and {Kawabata}, Miho and {Alsaberi}, Rami Z.~E. and {Dukiya}, Naveen and {Teja}, Rishabh Singh and {Ailawadhi}, Bhavya and {Dutta}, Anirban and {Sahu}, D.~K. and {Moriya}, Takashi J. and {Misra}, Kuntal and {Tanaka}, Masaomi and {Chevalier}, Roger and {Tominaga}, Nozomu and {Uno}, Kohki and {Imazawa}, Ryo and {Hamada}, Taisei and {Hori}, Tomoya and {Isogai}, Keisuke},
        title = "{Bridging between Type IIb and Ib Supernovae: SN IIb 2022crv with a Very Thin Hydrogen Envelope}",
      journal = {\apj},
         year = 2023,
        month = nov,
       volume = {957},
       number = {2},
          eid = {100},
        pages = {100},
          doi = {10.3847/1538-4357/acfa94},
archivePrefix = {arXiv},
       eprint = {2309.07463},
 primaryClass = {astro-ph.HE},
       adsurl = {https://ui.adsabs.harvard.edu/abs/2023ApJ...957..100G}
}

@ARTICLE{Chandra2019,
       author = {{Chandra}, Poonam and {Nayana}, A.~J. and {Bj{\"o}rnsson}, C. -I. and {Taddia}, Francesco and {Lundqvist}, Peter and {Ray}, Alak K. and {Shappee}, Benjamin J.},
        title = "{Type Ib Supernova Master OT J120451.50+265946.6: Radio-emitting Shock with Inhomogeneities Crossing through a Dense Shell}",
      journal = {\apj},
         year = 2019,
        month = jun,
       volume = {877},
       number = {2},
          eid = {79},
        pages = {79},
          doi = {10.3847/1538-4357/ab1900},
archivePrefix = {arXiv},
       eprint = {1904.06392},
 primaryClass = {astro-ph.HE},
       adsurl = {https://ui.adsabs.harvard.edu/abs/2019ApJ...877...79C}
}

@ARTICLE{Maoz2014,
       author = {{Maoz}, Dan and {Mannucci}, Filippo and {Nelemans}, Gijs},
        title = "{Observational Clues to the Progenitors of Type Ia Supernovae}",
      journal = {\araa},
         year = 2014,
        month = aug,
       volume = {52},
        pages = {107-170},
          doi = {10.1146/annurev-astro-082812-141031},
archivePrefix = {arXiv},
       eprint = {1312.0628},
 primaryClass = {astro-ph.CO},
       adsurl = {https://ui.adsabs.harvard.edu/abs/2014ARA&A..52..107M}
}

@ARTICLE{2016ApJ...823..100H,
       author = {{Harris}, Chelsea E. and {Nugent}, Peter E. and {Kasen}, Daniel N.},
        title = "{Against the Wind: Radio Light Curves of Type Ia Supernovae Interacting with Low-density Circumstellar Shells}",
      journal = {\apj},
         year = 2016,
        month = jun,
       volume = {823},
       number = {2},
          eid = {100},
        pages = {100},
          doi = {10.3847/0004-637X/823/2/100},
archivePrefix = {arXiv},
       eprint = {1604.00019},
 primaryClass = {astro-ph.HE},
       adsurl = {https://ui.adsabs.harvard.edu/abs/2016ApJ...823..100H}
}

@ARTICLE{2025ApJ...995...54G,
       author = {{Griffith}, Olivia and {Showerman}, Grace and {Sarbadhicary}, Sumit K. and {Harris}, Chelsea E. and {Chomiuk}, Laura and {Sollerman}, Jesper and {Lundqvist}, Peter and {Mold{\'o}n}, Javier and {Torres}, Miguel P{\'e}rez- and {Kool}, Erik C. and {Moriya}, Takashi J.},
        title = "{A Late-time Radio Survey of Type Ia-CSM Supernovae with the Very Large Array}",
      journal = {\apj},
         year = 2025,
        month = dec,
       volume = {995},
       number = {1},
          eid = {54},
        pages = {54},
          doi = {10.3847/1538-4357/ae17b0},
archivePrefix = {arXiv},
       eprint = {2506.19071},
 primaryClass = {astro-ph.HE},
       adsurl = {https://ui.adsabs.harvard.edu/abs/2025ApJ...995...54G}
}

@ARTICLE{2012ApJ...746...21H,
       author = {{Horesh}, Assaf and {Kulkarni}, S.~R. and {Fox}, Derek B. and {Carpenter}, John and {Kasliwal}, Mansi M. and {Ofek}, Eran O. and {Quimby}, Robert and {Gal-Yam}, Avishay and {Cenko}, S. Bradley and {de Bruyn}, A.~G. and {Kamble}, Atish and {Wijers}, Ralph A.~M.~J. and {van der Horst}, Alexander J. and {Kouveliotou}, Chryssa and {Podsiadlowski}, Philipp and {Sullivan}, Mark and {Maguire}, Kate and {Howell}, D. Andrew and {Nugent}, Peter E. and {Gehrels}, Neil and {Law}, Nicholas M. and {Poznanski}, Dovi and {Shara}, Michael},
        title = "{Early Radio and X-Ray Observations of the Youngest nearby Type Ia Supernova PTF 11kly (SN 2011fe)}",
      journal = {\apj},
         year = 2012,
        month = feb,
       volume = {746},
       number = {1},
          eid = {21},
        pages = {21},
          doi = {10.1088/0004-637X/746/1/21},
archivePrefix = {arXiv},
       eprint = {1109.2912},
 primaryClass = {astro-ph.CO},
       adsurl = {https://ui.adsabs.harvard.edu/abs/2012ApJ...746...21H}
}

@article{filippenko1997optical,
       author = {{Filippenko}, Alexei V.},
        title = "{Optical Spectra of Supernovae}",
      journal = {\araa},
         year = 1997,
        month = jan,
       volume = {35},
        pages = {309-355},
          doi = {10.1146/annurev.astro.35.1.309},
       adsurl = {https://ui.adsabs.harvard.edu/abs/1997ARA&A..35..309F}
}

@article{Hamuy2003_2002ic,
       author = {{Hamuy}, Mario and {Phillips}, M.~M. and {Suntzeff}, Nicholas B. and {Maza}, Jos{\'e} and {Gonz{\'a}lez}, L.~E. and {Roth}, Miguel and {Krisciunas}, Kevin and {Morrell}, Nidia and {Green}, E.~M. and {Persson}, S.~E. and {McCarthy}, P.~J.},
        title = "{An asymptotic-giant-branch star in the progenitor system of a type Ia supernova}",
      journal = {\nat},
         year = 2003,
        month = aug,
       volume = {424},
       number = {6949},
        pages = {651-654},
          doi = {10.1038/nature01854},
archivePrefix = {arXiv},
       eprint = {astro-ph/0306270},
 primaryClass = {astro-ph},
       adsurl = {https://ui.adsabs.harvard.edu/abs/2003Natur.424..651H}
}

@ARTICLE{Nayana2020,
       author = {{Nayana}, A.~J. and {Chandra}, Poonam},
        title = "{Radio view of a broad-line Type Ic supernova ASASSN-16fp}",
      journal = {\mnras},
         year = 2020,
        month = may,
       volume = {494},
       number = {1},
        pages = {84-96},
          doi = {10.1093/mnras/staa700},
archivePrefix = {arXiv},
       eprint = {2003.02709},
 primaryClass = {astro-ph.HE},
       adsurl = {https://ui.adsabs.harvard.edu/abs/2020MNRAS.494...84N}
}

@ARTICLE{Kulkarni1998,
       author = {{Kulkarni}, S.~R. and {Frail}, D.~A. and {Wieringa}, M.~H. and {Ekers}, R.~D. and {Sadler}, E.~M. and {Wark}, R.~M. and {Higdon}, J.~L. and {Phinney}, E.~S. and {Bloom}, J.~S.},
        title = "{Radio emission from the unusual supernova 1998bw and its association with the {\ensuremath{\gamma}}-ray burst of 25 April 1998}",
      journal = {\nat},
         year = 1998,
        month = oct,
       volume = {395},
       number = {6703},
        pages = {663-669},
          doi = {10.1038/27139},
       adsurl = {https://ui.adsabs.harvard.edu/abs/1998Natur.395..663K}
}

@ARTICLE{Nayana2018,
       author = {{Nayana}, A.~J. and {Chandra}, Poonam and {Ray}, Alak K.},
        title = "{Long-term Behavior of a Type IIP Supernova SN 2004dj in the Radio Bands}",
      journal = {\apj},
         year = 2018,
        month = aug,
       volume = {863},
       number = {2},
          eid = {163},
        pages = {163},
          doi = {10.3847/1538-4357/aad17a},
       adsurl = {https://ui.adsabs.harvard.edu/abs/2018ApJ...863..163N}
}

@ARTICLE{Weiler2011,
       author = {{Weiler}, Kurt W. and {Panagia}, Nino and {Stockdale}, Christopher and {Rupen}, Michael and {Sramek}, Richard A. and {Williams}, Christopher L.},
        title = "{Radio Emission from SN 1994I in NGC 5194 (M 51): The Best-studied Type Ib/c Radio Supernova}",
      journal = {\apj},
         year = 2011,
        month = oct,
       volume = {740},
       number = {2},
          eid = {79},
        pages = {79},
          doi = {10.1088/0004-637X/740/2/79},
archivePrefix = {arXiv},
       eprint = {1108.2692},
 primaryClass = {astro-ph.CO},
       adsurl = {https://ui.adsabs.harvard.edu/abs/2011ApJ...740...79W}
}

@incollection{Chevalier2003,
  author    = {Roger A. Chevalier and Claes Fransson},
  title     = {Supernova Interaction with a Circumstellar Medium},
  booktitle = {Supernovae and Gamma-Ray Bursters},
  editor    = {Kurt W. Weiler},
  series    = {Lecture Notes in Physics},
  volume    = {598},
  pages     = {171--194},
  year      = {2003},
  publisher = {Springer},
  address   = {Berlin, Heidelberg},
  doi       = {10.1007/3-540-45863-8_10}
}

@ARTICLE{Chandra2025,
       author = {{Chandra}, Poonam},
        title = "{Multiwavelength view of circumstellar interaction in supernovae}",
      journal = {arXiv e-prints},
         year = 2025,
        month = oct,
          eid = {arXiv:2510.20913},
        pages = {arXiv:2510.20913},
          doi = {10.48550/arXiv.2510.20913},
archivePrefix = {arXiv},
       eprint = {2510.20913},
 primaryClass = {astro-ph.HE},
       adsurl = {https://ui.adsabs.harvard.edu/abs/2025arXiv251020913C}
}

@ARTICLE{Smith2014,
       author = {{Smith}, Nathan},
        title = "{Mass Loss: Its Effect on the Evolution and Fate of High-Mass Stars}",
      journal = {\araa},
         year = 2014,
        month = aug,
       volume = {52},
        pages = {487-528},
          doi = {10.1146/annurev-astro-081913-040025},
archivePrefix = {arXiv},
       eprint = {1402.1237},
 primaryClass = {astro-ph.SR},
       adsurl = {https://ui.adsabs.harvard.edu/abs/2014ARA&A..52..487S}
}

@ARTICLE{Margutti2023,
       author = {{Margutti}, Raffaella and {Bright}, J.~S. and {Matthews}, D.~J. and {Coppejans}, D.~L. and {Alexander}, K.~D. and {Berger}, E. and {Bietenholz}, M. and {Chornock}, R. and {DeMarchi}, L. and {Drout}, M.~R. and {Eftekhari}, T. and {Jacobson-Gal{\'a}n}, W.~V. and {Laskar}, T. and {Milisavljevic}, D. and {Murase}, K. and {Nicholl}, M. and {Omand}, C.~M.~B. and {Stroh}, M. and {Terreran}, G. and {VanderLey}, B.~A.},
        title = "{Luminous Radio Emission from the Superluminous Supernova 2017ens at 3.3 yr after Explosion}",
      journal = {\apjl},
         year = 2023,
        month = sep,
       volume = {954},
       number = {2},
          eid = {L45},
        pages = {L45},
          doi = {10.3847/2041-8213/acf1fd},
archivePrefix = {arXiv},
       eprint = {2306.13730},
 primaryClass = {astro-ph.HE},
       adsurl = {https://ui.adsabs.harvard.edu/abs/2023ApJ...954L..45M}
}

@ARTICLE{Mondal2020,
       author = {{Mondal}, Surajit and {Bera}, Apurba and {Chandra}, Poonam and {Das}, Barnali},
        title = "{Radio emission from a decade old Type I superluminous supernova, PTF10hgi: comparison with FRB121102}",
      journal = {\mnras},
         year = 2020,
        month = nov,
       volume = {498},
       number = {3},
        pages = {3863-3869},
          doi = {10.1093/mnras/staa2637},
archivePrefix = {arXiv},
       eprint = {2008.11375},
 primaryClass = {astro-ph.HE},
       adsurl = {https://ui.adsabs.harvard.edu/abs/2020MNRAS.498.3863M}
}

@ARTICLE{Moriya2024,
       author = {{Moriya}, Takashi J.},
        title = "{Superluminous supernovae}",
      journal = {arXiv e-prints},
         year = 2024,
        month = jul,
          eid = {arXiv:2407.12302},
        pages = {arXiv:2407.12302},
          doi = {10.48550/arXiv.2407.12302},
archivePrefix = {arXiv},
       eprint = {2407.12302},
 primaryClass = {astro-ph.HE},
       adsurl = {https://ui.adsabs.harvard.edu/abs/2024arXiv240712302M}
}

@article{Chevalier1998,
       author = {{Chevalier}, Roger A.},
        title = "{Synchrotron Self-Absorption in Radio Supernovae}",
      journal = {\apj},
         year = 1998,
        month = may,
       volume = {499},
       number = {2},
        pages = {810-819},
          doi = {10.1086/305676},
       adsurl = {https://ui.adsabs.harvard.edu/abs/1998ApJ...499..810C}
}

@ARTICLE{Kool2023,
       author = {{Kool}, Erik C. and {Johansson}, Joel and {Sollerman}, Jesper and {Mold{\'o}n}, Javier and {Moriya}, Takashi J. and {Mattila}, Seppo and {Schulze}, Steve and {Chomiuk}, Laura and {P{\'e}rez-Torres}, Miguel and {Harris}, Chelsea and {Lundqvist}, Peter and {Graham}, Matthew and {Yang}, Sheng and {Perley}, Daniel A. and {Strotjohann}, Nora Linn and {Fremling}, Christoffer and {Gal-Yam}, Avishay and {Lezmy}, Jeremy and {Maguire}, Kate and {Omand}, Conor and {Smith}, Mathew and {Andreoni}, Igor and {Bellm}, Eric C. and {Bloom}, Joshua S. and {De}, Kishalay and {Groom}, Steven L. and {Kasliwal}, Mansi M. and {Masci}, Frank J. and {Medford}, Michael S. and {Park}, Sungmin and {Purdum}, Josiah and {Reynolds}, Thomas M. and {Riddle}, Reed and {Robert}, Estelle and {Ryder}, Stuart D. and {Sharma}, Yashvi and {Stern}, Daniel},
        title = "{A radio-detected type Ia supernova with helium-rich circumstellar material}",
      journal = {\nat},
         year = 2023,
        month = may,
       volume = {617},
       number = {7961},
        pages = {477-482},
          doi = {10.1038/s41586-023-05916-w},
archivePrefix = {arXiv},
       eprint = {2210.07725},
 primaryClass = {astro-ph.HE},
       adsurl = {https://ui.adsabs.harvard.edu/abs/2023Natur.617..477K}
}

@ARTICLE{Bietenholz2021,
       author = {{Bietenholz}, M.~F. and {Bartel}, N. and {Argo}, M. and {Dua}, R. and {Ryder}, S. and {Soderberg}, A.},
        title = "{The Radio Luminosity-risetime Function of Core-collapse Supernovae}",
      journal = {\apj},
         year = 2021,
        month = feb,
       volume = {908},
       number = {1},
          eid = {75},
        pages = {75},
          doi = {10.3847/1538-4357/abccd9},
archivePrefix = {arXiv},
       eprint = {2011.11737},
 primaryClass = {astro-ph.HE},
       adsurl = {https://ui.adsabs.harvard.edu/abs/2021ApJ...908...75B}
}

@ARTICLE{Petropoulou2016,
       author = {{Petropoulou}, M. and {Kamble}, A. and {Sironi}, L.},
        title = "{Radio synchrotron emission from secondary electrons in interaction-powered supernovae}",
      journal = {\mnras},
         year = 2016,
        month = jul,
       volume = {460},
       number = {1},
        pages = {44-66},
          doi = {10.1093/mnras/stw920},
archivePrefix = {arXiv},
       eprint = {1603.00891},
 primaryClass = {astro-ph.HE},
       adsurl = {https://ui.adsabs.harvard.edu/abs/2016MNRAS.460...44P}
}

@INPROCEEDINGS{BenAmi2022,
       author = {{Ben-Ami}, Sagi and {Shvartzvald}, Yossi and {Waxman}, Eli and {Netzer}, Udi and {Yaniv}, Yoram and {Algranatti}, Viktor M. and {Gal-Yam}, Avishay and {Lapid}, Ofer and {Ofek}, Eran and {Topaz}, Jeremy and {Arcavi}, Iair and {Asif}, Arooj and {Azaria}, Shlomi and {Bahalul}, Eran and {Barschke}, Merlin F. and {Bastian-Querner}, Benjamin and {Berge}, David and {Berlea}, Vlad D. and {Buehler}, Rolf and {Dittmar}, Louise and {Gelman}, Anatoly and {Giavitto}, Gianluca and {Guttman}, Or and {Haces Crespo}, Juan M. and {Heilbrunn}, Daniel and {Kachergincky}, Arik and {Kaipachery}, Nirmal and {Kowalski}, Marek and {Kulkarni}, Shrinivasrao R. and {Kumar}, Shashank and {K{\"u}sters}, Daniel and {Liran}, Tuvia and {Miron-Salomon}, Yonit and {Mor}, Zohar and {Nir}, Aharon and {Nitzan}, Gadi and {Philipp}, Sebastian and {Porelli}, Andrea and {Sagiv}, Ilan and {Schliwinski}, Julian and {Sprecher}, Tuvia and {De Simone}, Nicola and {Stern}, Nir and {Stone}, Nicholas C. and {Trakhtenbrot}, Benny and {Vasilev}, Mikhail and {Watson}, Jason J. and {Zappon}, Francesco},
        title = "{The scientific payload of the Ultraviolet Transient Astronomy Satellite (ULTRASAT)}",
    booktitle = {Space Telescopes and Instrumentation 2022: Ultraviolet to Gamma Ray},
         year = 2022,
       editor = {{den Herder}, Jan-Willem A. and {Nikzad}, Shouleh and {Nakazawa}, Kazuhiro},
       series = {Society of Photo-Optical Instrumentation Engineers (SPIE) Conference Series},
       volume = {12181},
        month = aug,
          eid = {1218105},
        pages = {1218105},
          doi = {10.1117/12.2629850},
archivePrefix = {arXiv},
       eprint = {2208.00159},
 primaryClass = {astro-ph.IM},
       adsurl = {https://ui.adsabs.harvard.edu/abs/2022SPIE12181E..05B}
}

@article{Chevalier1982b,
       author = {{Chevalier}, R.~A.},
        title = "{The radio and X-ray emission from type II supernovae.}",
      journal = {\apj},
         year = 1982,
        month = aug,
       volume = {259},
        pages = {302-310},
          doi = {10.1086/160167},
       adsurl = {https://ui.adsabs.harvard.edu/abs/1982ApJ...259..302C}
}

@article{Chugai2007,
       author = {{Chugai}, Nikolai N. and {Chevalier}, Roger A. and {Utrobin}, Victor P.},
        title = "{Optical Signatures of Circumstellar Interaction in Type IIP Supernovae}",
      journal = {\apj},
         year = 2007,
        month = jun,
       volume = {662},
       number = {2},
        pages = {1136-1147},
          doi = {10.1086/518160},
archivePrefix = {arXiv},
       eprint = {astro-ph/0703468},
 primaryClass = {astro-ph},
       adsurl = {https://ui.adsabs.harvard.edu/abs/2007ApJ...662.1136C}
}

@ARTICLE{Puls2008,
       author = {{Puls}, Joachim and {Vink}, Jorick S. and {Najarro}, Francisco},
        title = "{Mass loss from hot massive stars}",
      journal = {\aapr},
         year = 2008,
        month = dec,
       volume = {16},
       number = {3-4},
        pages = {209-325},
          doi = {10.1007/s00159-008-0015-8},
archivePrefix = {arXiv},
       eprint = {0811.0487},
 primaryClass = {astro-ph},
       adsurl = {https://ui.adsabs.harvard.edu/abs/2008A&ARv..16..209P}
}

@article{Woosley2002,
       author = {{Woosley}, S.~E. and {Heger}, A. and {Weaver}, T.~A.},
        title = "{The evolution and explosion of massive stars}",
      journal = {Reviews of Modern Physics},
         year = 2002,
        month = nov,
       volume = {74},
       number = {4},
        pages = {1015-1071},
          doi = {10.1103/RevModPhys.74.1015},
       adsurl = {https://ui.adsabs.harvard.edu/abs/2002RvMP...74.1015W}
}

@ARTICLE{Nomoto1982,
       author = {{Nomoto}, K.},
        title = "{Accreting white dwarf models for type I supernovae. I - Presupernova evolution and triggering mechanisms}",
      journal = {\apj},
         year = 1982,
        month = feb,
       volume = {253},
        pages = {798-810},
          doi = {10.1086/159682},
       adsurl = {https://ui.adsabs.harvard.edu/abs/1982ApJ...253..798N}
}

@article{WhelanIben1973,
       author = {{Whelan}, John and {Iben}, Jr., Icko},
        title = "{Binaries and Supernovae of Type I}",
      journal = {\apj},
         year = 1973,
        month = dec,
       volume = {186},
        pages = {1007-1014},
          doi = {10.1086/152565},
       adsurl = {https://ui.adsabs.harvard.edu/abs/1973ApJ...186.1007W}
}

@ARTICLE{Stroh2021,
       author = {{Stroh}, Michael C. and {Terreran}, Giacomo and {Coppejans}, Deanne L. and {Bright}, Joe S. and {Margutti}, Raffaella and {Bietenholz}, Michael F. and {De Colle}, Fabio and {DeMarchi}, Lindsay and {Duran}, Rodolfo Barniol and {Milisavljevic}, Danny and {Murase}, Kohta and {Paterson}, Kerry and {Williams}, Wendy L.},
        title = "{Luminous Late-time Radio Emission from Supernovae Detected by the Karl G. Jansky Very Large Array Sky Survey (VLASS)}",
      journal = {\apjl},
         year = 2021,
        month = dec,
       volume = {923},
       number = {2},
          eid = {L24},
        pages = {L24},
          doi = {10.3847/2041-8213/ac375e},
archivePrefix = {arXiv},
       eprint = {2106.09737},
 primaryClass = {astro-ph.HE},
       adsurl = {https://ui.adsabs.harvard.edu/abs/2021ApJ...923L..24S}
}

@ARTICLE{Chevalier2006,
       author = {{Chevalier}, Roger A. and {Fransson}, Claes},
        title = "{Circumstellar Emission from Type Ib and Ic Supernovae}",
      journal = {\apj},
         year = 2006,
        month = nov,
       volume = {651},
       number = {1},
        pages = {381-391},
          doi = {10.1086/507606},
archivePrefix = {arXiv},
       eprint = {astro-ph/0607196},
 primaryClass = {astro-ph},
       adsurl = {https://ui.adsabs.harvard.edu/abs/2006ApJ...651..381C}
}

@ARTICLE{Granot2002,
       author = {{Granot}, Jonathan and {Panaitescu}, Alin and {Kumar}, Pawan and {Woosley}, Stan E.},
        title = "{Off-Axis Afterglow Emission from Jetted Gamma-Ray Bursts}",
      journal = {\apjl},
         year = 2002,
        month = may,
       volume = {570},
       number = {2},
        pages = {L61-L64},
          doi = {10.1086/340991},
archivePrefix = {arXiv},
       eprint = {astro-ph/0201322},
 primaryClass = {astro-ph},
       adsurl = {https://ui.adsabs.harvard.edu/abs/2002ApJ...570L..61G}
}

@INCOLLECTION{Slane2017,
       author = {{Slane}, Patrick},
        title = "{Pulsar Wind Nebulae}",
    booktitle = {Handbook of Supernovae},
         year = 2017,
       editor = {{Alsabti}, Athem W. and {Murdin}, Paul},
        pages = {2159},
          doi = {10.1007/978-3-319-21846-5_95},
       adsurl = {https://ui.adsabs.harvard.edu/abs/2017hsn..book.2159S}
}

@ARTICLE{Nayana2025,
       author = {{Nayana}, A.~J. and {Margutti}, Raffaella and {Wiston}, Eli and {Chornock}, Ryan and {Campana}, Sergio and {Laskar}, Tanmoy and {Murase}, Kohta and {Krips}, Melanie and {Migliori}, Giulia and {Tsuna}, Daichi and {Alexander}, Kate D. and {Chandra}, Poonam and {Bietenholz}, Michael and {Berger}, Edo and {Chevalier}, Roger A. and {De Colle}, Fabio and {Dessart}, Luc and {Diesing}, Rebecca and {Grefenstette}, Brian W. and {Jacobson-Gal{\'a}n}, Wynn V. and {Maeda}, Keiichi and {Marcote}, Benito and {Matthews}, Daisy and {Milisavljevic}, Dan and {Ray}, Alak K. and {Reguitti}, Andrea and {Polzin}, Ava},
        title = "{Dinosaur in a Haystack: X-Ray View of the Entrails of SN 2023ixf and the Radio Afterglow of Its Interaction with the Medium Spawned by the Progenitor Star (Paper I)}",
      journal = {\apj},
         year = 2025,
        month = may,
       volume = {985},
       number = {1},
          eid = {51},
        pages = {51},
          doi = {10.3847/1538-4357/adc2fb},
archivePrefix = {arXiv},
       eprint = {2411.02647},
 primaryClass = {astro-ph.HE},
       adsurl = {https://ui.adsabs.harvard.edu/abs/2025ApJ...985...51N}
}

@ARTICLE{Murase2014,
       author = {{Murase}, Kohta and {Thompson}, Todd A. and {Ofek}, Eran O.},
        title = "{Probing cosmic ray ion acceleration with radio-submm and gamma-ray emission from interaction-powered supernovae}",
      journal = {\mnras},
         year = 2014,
        month = may,
       volume = {440},
       number = {3},
        pages = {2528-2543},
          doi = {10.1093/mnras/stu384},
archivePrefix = {arXiv},
       eprint = {1311.6778},
 primaryClass = {astro-ph.HE},
       adsurl = {https://ui.adsabs.harvard.edu/abs/2014MNRAS.440.2528M}
}

@ARTICLE{Murase2011,
       author = {{Murase}, Kohta and {Thompson}, Todd A. and {Lacki}, Brian C. and {Beacom}, John F.},
        title = "{New class of high-energy transients from crashes of supernova ejecta with massive circumstellar material shells}",
      journal = {\prd},
         year = 2011,
        month = aug,
       volume = {84},
       number = {4},
          eid = {043003},
        pages = {043003},
          doi = {10.1103/PhysRevD.84.043003},
archivePrefix = {arXiv},
       eprint = {1012.2834},
 primaryClass = {astro-ph.HE},
       adsurl = {https://ui.adsabs.harvard.edu/abs/2011PhRvD..84d3003M}
}

@INPROCEEDINGS{Katz2012,
       author = {{Katz}, Boaz and {Sapir}, Nir and {Waxman}, Eli},
        title = "{X-rays, {\ensuremath{\gamma}}-rays and neutrinos from collisionless shocks in supernova wind breakouts}",
    booktitle = {Death of Massive Stars: Supernovae and Gamma-Ray Bursts},
         year = 2012,
       editor = {{Roming}, P. and {Kawai}, N. and {Pian}, E.},
       series = {IAU Symposium},
       volume = {279},
        month = sep,
        pages = {274-281},
          doi = {10.1017/S174392131201304X},
archivePrefix = {arXiv},
       eprint = {1106.1898},
 primaryClass = {astro-ph.HE},
       adsurl = {https://ui.adsabs.harvard.edu/abs/2012IAUS..279..274K}
}

@ARTICLE{Kashiyama2013,
       author = {{Kashiyama}, Kazumi and {Murase}, Kohta and {Horiuchi}, Shunsaku and {Gao}, Shan and {M{\'e}sz{\'a}ros}, Peter},
        title = "{High-energy Neutrino and Gamma-Ray Transients from Trans-relativistic Supernova Shock Breakouts}",
      journal = {\apjl},
         year = 2013,
        month = may,
       volume = {769},
       number = {1},
          eid = {L6},
        pages = {L6},
          doi = {10.1088/2041-8205/769/1/L6},
archivePrefix = {arXiv},
       eprint = {1210.8147},
 primaryClass = {astro-ph.HE},
       adsurl = {https://ui.adsabs.harvard.edu/abs/2013ApJ...769L...6K}
}

@ARTICLE{Falk1973,
       author = {{Falk}, Sydney W. and {Arnett}, W. David},
        title = "{A Theoretical Model for Type II Supernovae}",
      journal = {\apjl},
         year = 1973,
        month = mar,
       volume = {180},
        pages = {L65},
          doi = {10.1086/181154},
       adsurl = {https://ui.adsabs.harvard.edu/abs/1973ApJ...180L..65F}
}

@ARTICLE{Ofek2007,
       author = {{Ofek}, E.~O. and {Cameron}, P.~B. and {Kasliwal}, M.~M. and {Gal-Yam}, A. and {Rau}, A. and {Kulkarni}, S.~R. and {Frail}, D.~A. and {Chandra}, P. and {Cenko}, S.~B. and {Soderberg}, A.~M. and {Immler}, S.},
        title = "{SN 2006gy: An Extremely Luminous Supernova in the Galaxy NGC 1260}",
      journal = {\apjl},
         year = 2007,
        month = apr,
       volume = {659},
       number = {1},
        pages = {L13-L16},
          doi = {10.1086/516749},
archivePrefix = {arXiv},
       eprint = {astro-ph/0612408},
 primaryClass = {astro-ph},
       adsurl = {https://ui.adsabs.harvard.edu/abs/2007ApJ...659L..13O}
}

@ARTICLE{Smith2007,
       author = {{Smith}, Nathan and {McCray}, Richard},
        title = "{Shell-shocked Diffusion Model for the Light Curve of SN 2006gy}",
      journal = {\apjl},
         year = 2007,
        month = dec,
       volume = {671},
       number = {1},
        pages = {L17-L20},
          doi = {10.1086/524681},
archivePrefix = {arXiv},
       eprint = {0710.3428},
 primaryClass = {astro-ph},
       adsurl = {https://ui.adsabs.harvard.edu/abs/2007ApJ...671L..17S}
}

@ARTICLE{Quimby2011,
       author = {{Quimby}, R.~M. and {Kulkarni}, S.~R. and {Kasliwal}, M.~M. and {Gal-Yam}, A. and {Arcavi}, I. and {Sullivan}, M. and {Nugent}, P. and {Thomas}, R. and {Howell}, D.~A. and {Nakar}, E. and {Bildsten}, L. and {Theissen}, C. and {Law}, N.~M. and {Dekany}, R. and {Rahmer}, G. and {Hale}, D. and {Smith}, R. and {Ofek}, E.~O. and {Zolkower}, J. and {Velur}, V. and {Walters}, R. and {Henning}, J. and {Bui}, K. and {McKenna}, D. and {Poznanski}, D. and {Cenko}, S.~B. and {Levitan}, D.},
        title = "{Hydrogen-poor superluminous stellar explosions}",
      journal = {\nat},
         year = 2011,
        month = jun,
       volume = {474},
       number = {7352},
        pages = {487-489},
          doi = {10.1038/nature10095},
archivePrefix = {arXiv},
       eprint = {0910.0059},
 primaryClass = {astro-ph.CO},
       adsurl = {https://ui.adsabs.harvard.edu/abs/2011Natur.474..487Q}
}

@ARTICLE{Yaron2017,
       author = {{Yaron}, O. and {Perley}, D.~A. and {Gal-Yam}, A. and {Groh}, J.~H. and {Horesh}, A. and {Ofek}, E.~O. and {Kulkarni}, S.~R. and {Sollerman}, J. and {Fransson}, C. and {Rubin}, A. and {Szabo}, P. and {Sapir}, N. and {Taddia}, F. and {Cenko}, S.~B. and {Valenti}, S. and {Arcavi}, I. and {Howell}, D.~A. and {Kasliwal}, M.~M. and {Vreeswijk}, P.~M. and {Khazov}, D. and {Fox}, O.~D. and {Cao}, Y. and {Gnat}, O. and {Kelly}, P.~L. and {Nugent}, P.~E. and {Filippenko}, A.~V. and {Laher}, R.~R. and {Wozniak}, P.~R. and {Lee}, W.~H. and {Rebbapragada}, U.~D. and {Maguire}, K. and {Sullivan}, M. and {Soumagnac}, M.~T.},
        title = "{Confined dense circumstellar material surrounding a regular type II supernova}",
      journal = {Nature Physics},
         year = 2017,
        month = feb,
       volume = {13},
       number = {5},
        pages = {510-517},
          doi = {10.1038/nphys4025},
archivePrefix = {arXiv},
       eprint = {1701.02596},
 primaryClass = {astro-ph.HE},
       adsurl = {https://ui.adsabs.harvard.edu/abs/2017NatPh..13..510Y}
}

@ARTICLE{Goldman2017,
       author = {{Goldman}, Steven R. and {van Loon}, Jacco Th. and {Zijlstra}, Albert A. and {Green}, James A. and {Wood}, Peter R. and {Nanni}, Ambra and {Imai}, Hiroshi and {Whitelock}, Patricia A. and {Matsuura}, Mikako and {Groenewegen}, Martin A.~T. and {G{\'o}mez}, Jos{\'e} F.},
        title = "{The wind speeds, dust content, and mass-loss rates of evolved AGB and RSG stars at varying metallicity}",
      journal = {\mnras},
         year = 2017,
        month = feb,
       volume = {465},
       number = {1},
        pages = {403-433},
          doi = {10.1093/mnras/stw2708},
archivePrefix = {arXiv},
       eprint = {1610.05761},
 primaryClass = {astro-ph.SR},
       adsurl = {https://ui.adsabs.harvard.edu/abs/2017MNRAS.465..403G}
}

@ARTICLE{Forster2018,
       author = {{F{\"o}rster}, F. and {Moriya}, T.~J. and {Maureira}, J.~C. and {Anderson}, J.~P. and {Blinnikov}, S. and {Bufano}, F. and {Cabrera-Vives}, G. and {Clocchiatti}, A. and {de Jaeger}, T. and {Est{\'e}vez}, P.~A. and {Galbany}, L. and {Gonz{\'a}lez-Gait{\'a}n}, S. and {Gr{\"a}fener}, G. and {Hamuy}, M. and {Hsiao}, E.~Y. and {Huentelemu}, P. and {Huijse}, P. and {Kuncarayakti}, H. and {Mart{\'\i}nez}, J. and {Medina}, G. and {Olivares E.}, F. and {Pignata}, G. and {Razza}, A. and {Reyes}, I. and {San Mart{\'\i}n}, J. and {Smith}, R.~C. and {Vera}, E. and {Vivas}, A.~K. and {de Ugarte Postigo}, A. and {Yoon}, S.-C. and {Ashall}, C. and {Fraser}, M. and {Gal-Yam}, A. and {Kankare}, E. and {Le Guillou}, L. and {Mazzali}, P.~A. and {Walton}, N.~A. and {Young}, D.~R.},
        title = "{The delay of shock breakout due to circumstellar material evident in most type II supernovae}",
      journal = {Nature Astronomy},
         year = 2018,
        month = sep,
       volume = {2},
        pages = {808},
          doi = {10.1038/s41550-018-0563-4},
archivePrefix = {arXiv},
       eprint = {1809.06379},
 primaryClass = {astro-ph.HE},
       adsurl = {https://ui.adsabs.harvard.edu/abs/2018NatAs...2..808F}
}

@ARTICLE{Rose2024,
       author = {{Rose}, Kovi and {Horesh}, Assaf and {Murphy}, Tara and {Kaplan}, David L. and {Sfaradi}, Itai and {Ryder}, Stuart D. and {Aloisi}, Robert J. and {Dobie}, Dougal and {Driessen}, Laura and {Fender}, Rob and {Green}, David A. and {Leung}, James K. and {Lenc}, Emil and {Qiu}, Hao and {Williams-Baldwin}, David},
        title = "{Late-time supernovae radio re-brightening in the VAST pilot survey}",
      journal = {\mnras},
         year = 2024,
        month = nov,
       volume = {534},
       number = {4},
        pages = {3853-3868},
          doi = {10.1093/mnras/stae2289},
archivePrefix = {arXiv},
       eprint = {2410.01375},
 primaryClass = {astro-ph.HE},
       adsurl = {https://ui.adsabs.harvard.edu/abs/2024MNRAS.534.3853R}
}

@ARTICLE{Murphy2013,
       author = {{Murphy}, Tara and {Chatterjee}, Shami and {Kaplan}, David L. and {Banyer}, Jay and {Bell}, Martin E. and {Bignall}, Hayley E. and {Bower}, Geoffrey C. and {Cameron}, Robert A. and {Coward}, David M. and {Cordes}, James M. and {Croft}, Steve and {Curran}, James R. and {Djorgovski}, S.~G. and {Farrell}, Sean A. and {Frail}, Dale A. and {Gaensler}, B.~M. and {Galloway}, Duncan K. and {Gendre}, Bruce and {Green}, Anne J. and {Hancock}, Paul J. and {Johnston}, Simon and {Kamble}, Atish and {Law}, Casey J. and {Lazio}, T. Joseph W. and {Lo}, Kitty K. and {Macquart}, Jean-Pierre and {Rea}, Nanda and {Rebbapragada}, Umaa and {Reynolds}, Cormac and {Ryder}, Stuart D. and {Schmidt}, Brian and {Soria}, Roberto and {Stairs}, Ingrid H. and {Tingay}, Steven J. and {Torkelsson}, Ulf and {Wagstaff}, Kiri and {Walker}, Mark and {Wayth}, Randall B. and {Williams}, Peter K.~G.},
        title = "{VAST: An ASKAP Survey for Variables and Slow Transients}",
      journal = {\pasa},
         year = 2013,
        month = feb,
       volume = {30},
          eid = {e006},
        pages = {e006},
          doi = {10.1017/pasa.2012.006},
archivePrefix = {arXiv},
       eprint = {1207.1528},
 primaryClass = {astro-ph.IM},
       adsurl = {https://ui.adsabs.harvard.edu/abs/2013PASA...30....6M}
}

@ARTICLE{Brunthaler2009,
       author = {{Brunthaler}, A. and {Menten}, K.~M. and {Reid}, M.~J. and {Henkel}, C. and {Bower}, G.~C. and {Falcke}, H.},
        title = "{Discovery of a bright radio transient in M 82: a new radio supernova?}",
      journal = {\aap},
         year = 2009,
        month = may,
       volume = {499},
       number = {2},
        pages = {L17-L20},
          doi = {10.1051/0004-6361/200912327},
archivePrefix = {arXiv},
       eprint = {0904.2388},
 primaryClass = {astro-ph.CO},
       adsurl = {https://ui.adsabs.harvard.edu/abs/2009A&A...499L..17B}
}

@ARTICLE{Weiler2010,
       author = {{Weiler}, K.~W. and {Panagia}, N. and {Sramek}, R.~A. and {Van Dyk}, S.~D. and {Stockdale}, C.~J. and {Williams}, C.~L.},
        title = "{Radio emission from supernovae.}",
      journal = {\memsai},
         year = 2010,
        month = jan,
       volume = {81},
        pages = {374},
       adsurl = {https://ui.adsabs.harvard.edu/abs/2010MmSAI..81..374W}
}
